\documentclass[fleqn,10pt]{SelfArx} 

\usepackage[english]{babel} 

\usepackage{lipsum} 
\usepackage{afterpage}

 \definecolor{color1}{RGB}{0,0,110} 
\definecolor{color2}{RGB}{0,20,80} 

\usepackage{hyperref} 
\hypersetup{hidelinks,colorlinks,breaklinks=true,urlcolor=color2,citecolor=color1,linkcolor=color1,bookmarksopen=false,pdftitle={Title},pdfauthor={Author}}

\JournalInfo{Uploaded to arXiv.org} 
\Archive{} 

\PaperTitle{Application of a bipolar nanopore as a sensor: rectification as an additional device function} 

\Authors{Eszter M\'adai\textsuperscript{1,2}, M\'onika Valisk\'o\textsuperscript{1},  Dezs\H{o} Boda\textsuperscript{1}*} 
\affiliation{\textsuperscript{1}\textit{Department of Physical Chemistry, University of Pannonia,  P. O. Box 158, H-8201 Veszpr\'em, Hungary}} 
\affiliation{\textsuperscript{2}\textit{Department of Material- and Geo-Sciences, Technische Universit\"{a}t  Darmstadt, Petersenstr.\ 23, D-64287 Darmstadt, Germany}} %
\affiliation{*\textbf{Corresponding author}: dezsoboda@gmail.com} 

\newcommand{\keywordname}{Keywords} 

\Abstract{We model and simulate a nanopore sensor that selectively binds analyte ions.
This binding leads to the modulation of the local concentrations of the ions of the background electrolyte (KCl), and, thus, to the modulation of the ionic current flowing through the pore. 
The nanopore's wall carries a bipolar charge pattern with a larger positive buffer region determining the anions as the main charge carriers and the smaller negative binding region containing binding sites.
This charge pattern proved to be an appropriate one as shown by a previous comparative study of varying charge patterns (M\'adai et al.\ \textit{J. Mol. Liq.}, 2019, \textbf{283}, 391--398.).  
Binding of the positive analyte ions attracts more anions in the pore thus increasing the current.
The asymmetric nature of the pore results in an additional device function, rectification. 
Our model, therefore, is a dual response device.
Using a reduced model of the nanopore studied by a hybrid computer simulation method (Local Equilibrium Monte Carlo coupled to the Nernst-Planck equation) we show that we can create a sensor whose underlying mechanisms are based on the changes of the local electric field as a response to changing thermodynamic conditions. 
The change of the  electric field results in changes in the local ionic concentrations (depletion zones), and, thus, changes in ionic currents.}

\begin{document}

\flushbottom 

\maketitle 


\thispagestyle{empty} 



\section{Introduction}
\label{sec:intro}

The small diameter of nanopores makes it possible to use them as label-free sensors from the simple reason that molecules specifically binding to the functionalized wall of the pore can cause detectable change in the ionic current flowing through the pore.
According to thermodynamics, the probability that the pore captures an analyte molecule is related to its concentration in the bath.
This relation between the electrical read-out of the device and the analyte concentration makes it possible to design and fabricate efficient nanoscale sensors that can detect the analyte molecules even if they are present in very low concentrations.

In this paper, we study a nanopore-based sensor with an asymmetric charge pattern on the wall of the pore.
The left 2/3 of the pore is a buffer region whose charge determines the main charge carrier of the nanopore, while the right 1/3 of the pore is a region where analyte ions (denoted by X$^{z_{\mathrm{X}}}$ with $z_{\mathrm{X}}$ being the valence of the analyte ion; we consider only positively charged X ions) are bound by binding sites modeled with the square-well (SW) potential here.
This work is a direct continuation of our previous paper \cite{madai_jml_2019a} in which we allowed different charge densities (positive, negative, or zero) in the two regions and studied  the effect of varying charge patterns on the applicability of the nanopore as a sensor.
Surface charge pattern is a powerful tunable structural feature because it can be manipulated with chemical methods relatively easily \cite{stein_prl_2004,Siwy_2004,miedema_nl_2007,singh_jap_2011,nasir_acsami_2014,zhang_acsnano_2015}.

Two distinct charge patterns emerged that resulted in efficient sensors.
In one case, the whole pore is negatively charged, the charge carriers are the cations (K$^{+}$), and the sensor works on the basis of a competition between the background cations (K$^{+}$) and the analyte cations (X). 
If the bath concentration of the analyte ions, $c_{\mathrm{X}}$, is larger, more of them are bound at the binding sites, therefore, more K$^{+}$ ions will be replaced in the pore by X ions thus decreasing the K$^{+}$ current.
The relation of $c_{\mathrm{X}}$ and current is called calibration curve and it is the basis of the quantitative determination of the analyte concentration. 

This case was already studied in our first paper on the subject \cite{madai_jcp_2017} considering a symmetric sensor.
The device function is the current in the presence of the X ions ($I$) in relation to the current in the absence of the X ions ($I_{0}$).
This $I/I_{0}$ ratio is the only device function in this model.
Its value is a characteristic response of the device to changing input conditions: the presence of analyte ions at a given bath concentration, in this case.

If the charge pattern and/or the geometry of the nanopore is asymmetric, rectification appears as an additional device function \cite{siwy_nim_2003,hou_advmat_2010,cervera_ea_2011,zhang_cc_2013,ali_acsami_2015,zhang_csr_2018}.
Rectification is defined as the $|I(U)/I(-U)|$ ratio, where the sign of the voltage is defined so that $I(U)>|I(-U)|$.
In this case, we obtain a dual response device that has two device functions: the $I/I_{0}$ ratio and the rectification.

In our previous work \cite{madai_jml_2019a}, we established that the most efficient asymmetric charge pattern from the point of view of the  efficiency of sensing is the bipolar one where the left region is positively, while the right region is negatively charged.
In this case, the main charge carrier is the anion (Cl$^{-}$) and the sensor works on the basis of the increasing Cl$^{-}$ current caused by the increased accumulation of Cl$^{-}$ attracted by the positive analyte ions into the binding region.
There are well-established techniques for the fabrication of bipolar pores from chemical treatment to atomic layer deposition \cite{vlassiouk_nl_2007,vlassiouk_jacs_2009,wu_nanoscale_2012,sheng_bmf_2014}.

Our previous work reported results only for a restricted set of parameters (monovalent X$^{+}$, $c_{\mathrm{KCl}}=0.01$ M, cylindrical pore), while we explore an extended parameter space in this study with divalent and trivalent analyte ions, varying KCl concentrations, and conical nanopores.

Our study was inspired by experimental works. 
Several pairs of the analyte ions and the molecules that selectively bind them have been proposed in the literature.
The analyte ions can be Li$^{+}$ \cite{Ali_AC_2018}, Cs$^{+}$ \cite{Ali_Lang_2017}, Ca$^{2+}$/Mg$^{2+}$ \cite{Ali_ACSnano_2012}, K$^{+}$ \cite{liu_JACS_2015,wu_langmuir_2017,acar_sa_2019}, Na$^{+}$ \cite{liu_JACS_2015},  F$^{-}$  \cite{nie_chemsci_2015}, Zn$^{2+}$ \cite{tian_chemcomm_2010}, Cu$^{2+}$ \cite{zhao_ac_2017,deng_bsbe_2015}, Cd$^{2+}$ \cite{elsafty_afm_2007}, Hg$^{2+}$ \cite{mayne_acsnano_2018}, Pb$^{2+}$ \cite{mayne_acsnano_2018}, amino acids \cite{ali_cc_2015,vlassiouk_jacs_2009}, or sugars \cite{sun_chemcomm_2012}.
The molecules that bind them can be amine terminated acyclic polyether derivatives \cite{Ali_AC_2018}, calixcrown moieties \cite{Ali_Lang_2017,nie_chemsci_2015}, carboxylic acid and phosphonic polyacid chains \cite{Ali_ACSnano_2012}, crown ethers \cite{liu_JACS_2015,wu_langmuir_2017,acar_sa_2019}, zinc finger peptides \cite{tian_chemcomm_2010}, polyglutamic acids  \cite{zhao_ac_2017}, chromofore molecules synthesized to form nanosized cavities \cite{elsafty_afm_2007}, or aptamers \cite{mayne_acsnano_2018}.  
Even this short list shows the huge variety of the possibilities for specific binding that can be the basis of sensing. 
When the analyte is a biomolecule, the possibility for finding the appropriate binding counterpart (generally, another biomolecule) is especially large regarding that nature has used these binding pairs in living systems for a couple of billions of years.
These binding biomolecules can be enzymes \cite{ali_analchem_2011,hou_materchemA_2014,perezmitta_nl_2018}, antibodies\cite{vlassiouk_jacs_2009}, or DNA aptamers \cite{ali_cc_2015,mayne_acsnano_2018} just to mention a few.

There are various devices with which we can get a detectable signal resulting from these binding events.
Ensinger et al.\ \cite{Ali_AC_2018,Ali_Lang_2017,Ali_ACSnano_2012,liu_JACS_2015,ensinger_2018,nasir_jcis_2019} used conical polyethylene terephthalate (PET) nanopores that exhibit rectification due to their asymmetric geometries.
When metal ions are bound to the functionalized surfaces thus changing the surface charge pattern on the nanopores' wall, the conduction properties of the pore, including rectification, are changed.
When they exposed a negatively charged nanopore treated with immobilized DNA aptamer (LyzAp–NH$_{2}$) to lysozime (Lyz) protein only on one side, that side was made positive, and, thus, the pore was made bipolar \cite{ali_cc_2015}. 
It was shown that the rectification of this bipolar pore is sensitive to both Lyz and KCl concentrations. 

Vlassiouk et al.\ \cite{vlassiouk_jacs_2009} functionalized a conical nanopore with $\gamma$DPGA antibodies and, thus, created a pH-dependent charge asymmetry that superimposed the geometrical asymmetry. 
As a result of the pH-sensitive balance of the two kinds of asymmetries, an inversion of rectification as a function of pH was produced. 
Adding $\gamma$DPGA glutamic acids changed the charge pattern and modulated this balance. 
This is also a dual response channel sensitive to both pH and binding of analyte molecules. 

Zhao et al.\ have developed a ``nanochannel-ionchannel hybrid device'' that is an array of nanopores in porous anodic alumina (PAA) with channels in the barrier layer of PAA with diameters in the range  $0-0.8$ nm.
Polyglutamic acid (PGA) was used to chelate Cu$^{2+}$ resulting in a decrease of the effective cross sections of the pores \cite{zhao_ac_2017}.
In another work, a thrombin aptamer was used to bind thrombin and detect it in extremely low concentrations \cite{zhao_ac_2018}. 

Present paper is a modeling study using a simple implicit-water electrolyte model, where ions are modeled as charged hard spheres. 
As ionic sizes are relevant in the competitive mechanisms underlying selective sensing, we use a Monte Carlo (MC) method to treat the statistical mechanical problem.
Because we consider a non-equilibrium system, we use the Local Equilibrium Monte Carlo (LEMC) method \cite{boda_jctc_2012} that is an adaptation of the Grand Canonical Monte Carlo (GCMC) technique to a non-equilibrium situation by applying a space-dependent chemical potential profile.
The grand canonical nature of the computer simulation also makes it possible to simulate very low concentrations.
We consider concentrations as low as $10^{-10}$ M, in this study.

Because LEMC samples only the configurational degrees of freedom (ion positions), we need a method which can compute the flux.
We apply the Nernst-Planck (NP) transport equation and couple it to LEMC resulting in a hybrid method, called NP+LEMC.
This method was successfully applied for various problems in the last couple of years such as particle transport through model membranes \cite{boda_jctc_2012,hato_jcp_2012}, ion channels \cite{boda_jml_2014,boda_arcc_2014,hato_cmp_2016}, and nanopores \cite{hato_pccp_2017,matejczyk_jcp_2017,madai_jcp_2017,fertig_mp_2018,madai_pccp_2018,madai_jml_2019a,valisko_jcp_2019}.

\begin{figure*}[t]
	\begin{center}
		\includegraphics*[width=0.6\textwidth]{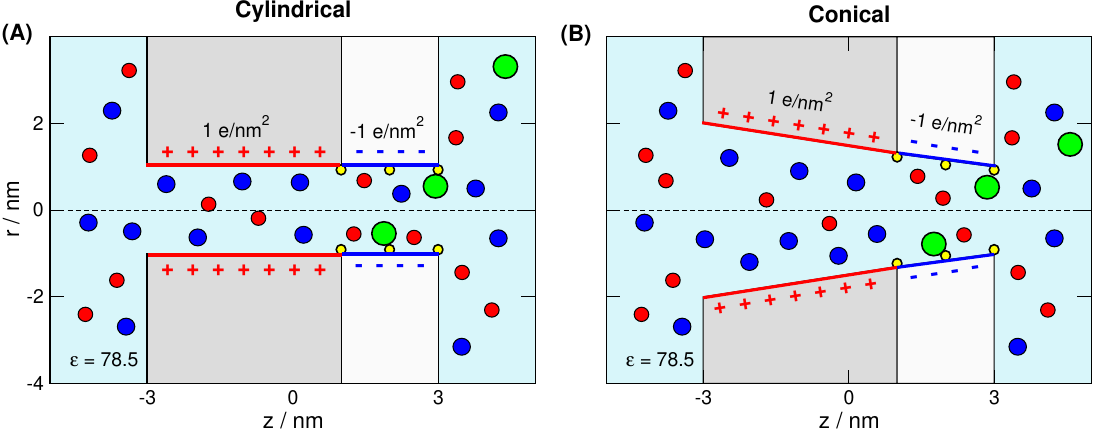}
	\end{center}
		\caption{Schematics of the pore that is divided into two regions. 
		The left region of length $4$ nm carries $1$ $e$/nm$^{2}$ surface charge that determines the main charge carrier of the pore, the anions, in this case.
		The right region is the ``binding region'', where the binding sites are located (indicated by small yellow spheres on the wall).
		The length of this region is $2$ nm and carries $-1$ $e$/nm$^{2}$ surface charge.
		We consider (A) a cylindrical pore with radius $1$ nm, and (B) a conical pore with $1$ nm radius at the tip (right entrance) and $2$ nm radius at the wide left entrance.
		The green sphere is the analyte ion, X$^{z_{\mathrm{X}}}$, that is bound to the binding site if it overlaps with the yellow sphere. 
		The red and blue spheres are the cations (K$^{+}$) and anions (Cl$^{-}$) of the electrolyte, respectively.
		} \label{fig1}
\end{figure*}

Reduced models proved to be useful many times in predicting relevant mechanisms behind device behavior \cite{eisenberg_jce_2002,eisenberg_jpcc_2010,eisenberg_jml_2018}.
Their success is counter-intuitive because these models ignore degrees of freedom that seem important on the molecular level such as explicit water molecules.
Reduced models work because they include those degrees of freedom that are relevant for device function.
This question has been analyzed in detail in our papers for bipolar nanopores \cite{hato_cmp_2016,hato_pccp_2017,valisko_jcp_2019}.

\section{Model and method}
\label{sec:model}

\subsection{Nanopore model}
\label{subsec:pore}

In this work, we consider two nanopores with different geometries.
One is a cylindrical pore of radius $1$ nm (Fig.\ \ref{fig1}A).
The other is a conical pore with radius $1$ nm at the tip on the right hand side entrance and radius $2$ nm at the wide entrance (base) of the left hand side (Fig.\ \ref{fig1}B).
Both pores penetrate a membrane of width $6$ nm, a value that defined pore length.
The walls of the pore and the membrane are hard, namely, overlap of ions with these walls is forbidden.

The pores are divided into two regions along the $z$-axis  ($z$ is the coordinate along the main axis of the pore, perpendicular to the membrane).
The left region of length $4$ nm  carries $1$ $e$/nm$^{2}$ surface charge.
This is a buffer region whose surface charge determines the main charge--carrier ionic species; the anions, in this particular case.
The right region of length $2$ nm, called binding region, contains the binding sites and carries $-1$ $e$/nm$^{2}$ surface charge.

The surface charge is represented by fractional point charges that are situated on a rectangular grid, where a surface element is approximately a square of size $0.2 \times 0.2$ nm$^{2}$.
The magnitude of the point charges is established so that the surface charge density corresponds to the prescribed values, $\pm 1$ $e$/nm$^{2}$.

Summarized, the radius at the right-hand-side tip, the lengths, and the axial surface charge patterns are common in the two nanopores, while they are different in the radii on the left hand side. 

\subsection{Interparticle potentials}

The ions of the electrolyte are modeled as charged hard spheres of charges, $q_{i}=z_{i}e$ ($z_{i}$ is the valence and $e$ is the elementary charge), and radii, $R_{i}$:
\begin{equation}
u_{ij}(r) =
\left\lbrace 
\begin{array}{ll}
\infty & \quad \mathrm{for} \quad r<R_{i}+R_{j} \\
 \dfrac{1}{4\pi\epsilon_{0}\epsilon} \dfrac{q_{i}q_{j}}{r} & \quad \mathrm{for} \quad r \geq R_{i}+R_{j}\\
\end{array}
\right. 
\label{eq:uij}
\end{equation} 
where $\epsilon_{0}$ is the permittivity of vacuum, $\epsilon=78.5$ is the dielectric constant of the electrolyte, and $r$ is the distance between two ions.
Subscript $i$ can take the values $i=\mathrm{K}^{+}$, $\mathrm{Cl}^{-}$, and X, where these symbols refer to the cation of the electolyte (K$^{+}$), the anion of the electrolyte (Cl$^{-}$), and the analyte ion (X$^{z_{\mathrm{X}}}$), respectively.
The valence of the analyte ion can be $z_{\mathrm{X}}=1$, $2$, and $3$ in this work.
We will refer to the analyte as just X for briefness most of the time.
The ionic radii are $R_{\mathrm{K}^{+}}=0.133$ nm, $R_{\mathrm{Cl}^{-}}=0.181$ nm (Pauling radii), and mostly $R_{\mathrm{X}}=0.3$ nm.
An analysis for $R_{\mathrm{X}}$--dependence is also reported.
The concentration of the KCl background electrolyte is varied in the range $c_{\mathrm{KCl}}=0.01-0.1$ M.

\begin{figure*}[t]
	\begin{center}
		\includegraphics*[width=0.6\textwidth]{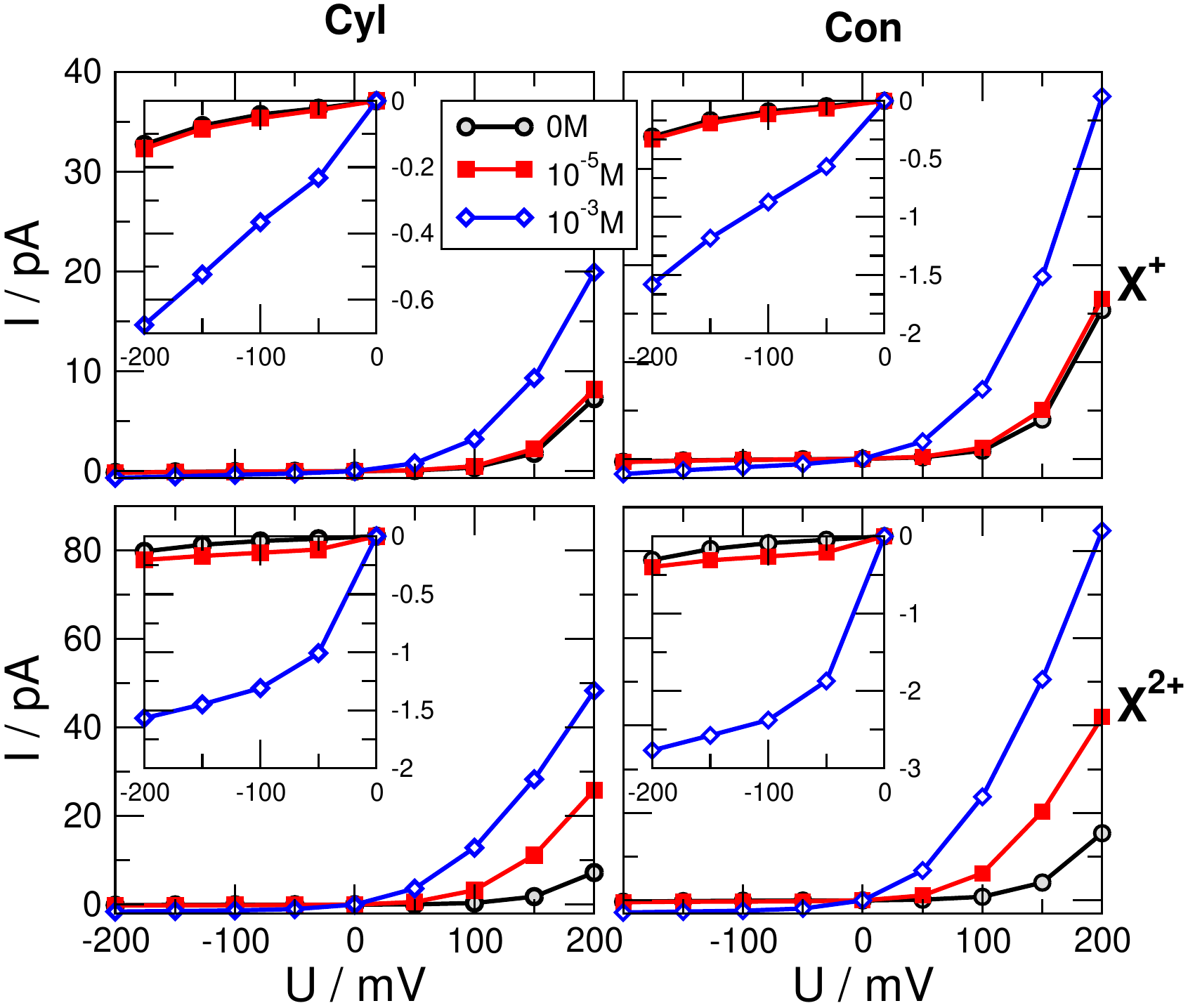}
		\caption{Current voltage curves for three different X concentrations, $0$ M (black circles), $10^{-5}$ M (red squares), and $10^{-3}$ M (blue diamonds) for $c_{\mathrm{KCl}}=0.01$ M background electrolyte concentration. 
		Left and right columns refer to cylindrical and conical geometries, respectively.
		Top and bottom rows refer to monovalent ($z_{\mathrm{X}}=1$) and divalent ($z_{\mathrm{X}}=2$) analyte ions, respectively.
		The insets zoom in on the data of the OFF state (negative voltages).} 
		\label{fig:fig2}
	\end{center}
\end{figure*}

The solvent is represented as a continuum background characterized by two response functions. 
One is the dielectric constant, $\epsilon$, that describes the screening effect of the water molecules.
The other is a diffusion coefficient function, $D_{i}(z)$, that describes the ability of water molecules to affect the diffusion of ions.
This function is space dependent in our case; it is a piecewise constant function that is different inside the pore ($D_{i}^{\mathrm{pore}}$)
and in the bulk ($D_{i}^{\mathrm{bulk}}$).
The bulk value is experimental ($D^{\mathrm{bulk}}_{\mathrm{K}^{+}}=1.849\times10^{-9}$ m$^{2}$s$^{-1}$ and $D^{\mathrm{bulk}}_{\mathrm{Cl}^{-}}=D^{\mathrm{bulk}}_{\mathrm{X}}=2.032\times10^{-9}$  m$^{2}$s$^{-1}$), while $D_{i}^{\mathrm{pore}}$ just scales the current without influencing the $I/I_{0}$ ratio.
Following our previous studies \cite{matejczyk_jcp_2017,madai_jcp_2017,fertig_mp_2018,madai_pccp_2018} here we set the relation $D_{i}^{\mathrm{pore}}=0.1\,D_{i}^{\mathrm{bulk}}$.
In other studies, where reference was available, we fitted the $D_{i}^{\mathrm{pore}}$ value to experimental \cite{boda_arcc_2014,fertig_mp_2018} or molecular dynamics \cite{hato_pccp_2017,valisko_jcp_2019} data.

We placed the binding sites on the pore wall in $3$ rings placed at $z=1$, $2$, and $3$ nm (yellow spheres in Fig.\ \ref{fig1}).
Each ring contains 4 binding sites \cite{madai_jcp_2017}.
The binding potential between a site and an analyte ion is the square-well (SW) potential:
\begin{equation}
u_{\mathrm{SW}}(r) =
\left\lbrace 
\begin{array}{ll}
0 & \quad \mathrm{for} \quad r-R_{\mathrm{X}}>d_{\mathrm{SW}}\\
 -\epsilon_{\mathrm{SW}} & \quad \mathrm{for} \quad r-R_{\mathrm{X}}<d_{\mathrm{SW}} ,
\end{array}
\right. 
\label{eq:squarewell}
\end{equation}
where $r$ is the distance of the site and the ion center.
This short-range potential attracts X with $-\epsilon_{\mathrm{SW}}=-10\,kT$ energy once the closest point of the X ion's surface is closer to the site than the distance parameter $d_{\mathrm{SW}}=0.2$ nm.
This model takes into account that the active site of the X ion is usually on its surface while keeping the spherical symmetry of the ion (it neglects the possible orientation dependence of binding).
The SW potential acts only on the X ions in the simulations. 
An analysis on the effect of the $\epsilon_{\mathrm{SW}} $ and $d_{\mathrm{SW}}$ parameters has been given in our previous work  \cite{madai_jcp_2017}.

\subsection{NP+LEMC}
\label{sec:method}

In the NP+LEMC  technique \cite{boda_jctc_2012} the NP equation,
\begin{equation}
 \mathbf{j}_{i}(\mathbf{r})=-\dfrac{1}{kT}D_{i}(\mathbf{r})c_{i}(\mathbf{r}) \nabla \mu_{i}(\mathbf{r}) ,
\label{eq:np}
\end{equation} 
is used to compute the particle flux density, $\mathbf{j}_{i}(\mathbf{r})$, for ion species $i$, where $T=298.15$ K is temperature, $k$ is Boltzmann's constant,  $c_{i}(\mathbf{r})$ is the concentration profile, and $\mu_{i}(\mathbf{r})$ is the electrochemical potential profile.

\begin{figure*}[t]
	\begin{center}
		\includegraphics*[width=0.6\textwidth]{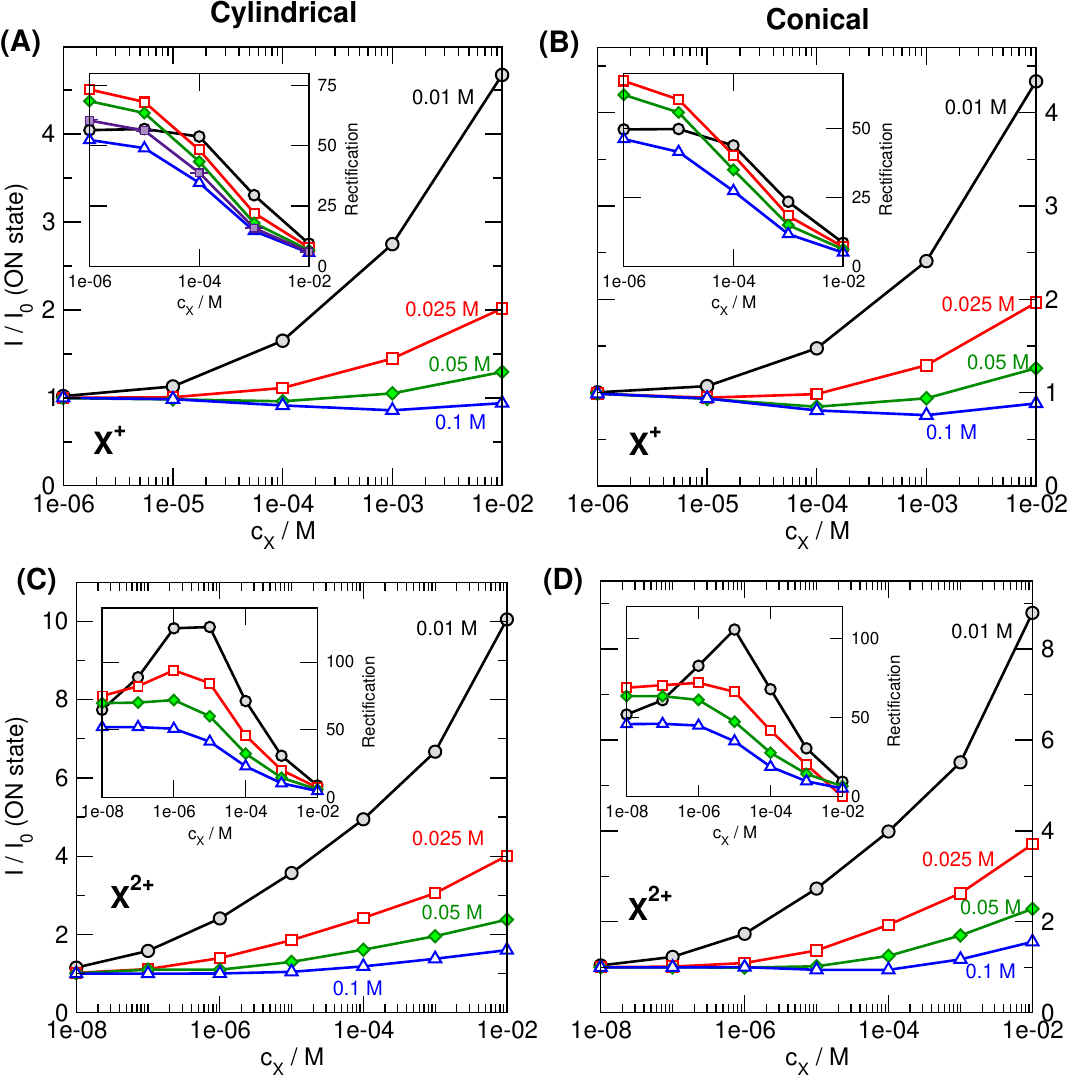}
		\caption{Relative currents ($I/I_{0}$ with $I_{0}$ being the current in the absence of X) in the ON state (200 mV) as functions of analyte concentration ($c_{\mathrm{X}}$). 
		The insets show the rectification, $I(200\mathrm{mV})/I(-200\mathrm{mV})$, as functions of $c_{\mathrm{X}}$.
		Left and right columns refer to cylindrical and conical geometries, respectively.
		Top and bottom rows refer to monovalent ($z_{\mathrm{X}}=1$) and divalent ($z_{\mathrm{X}}=2$) analyte ions, respectively.
		Various curves in a panel refer to different KCl concentrations indicated by the numbers near the curves.
		}
		\label{fig:fig3}
	\end{center}
\end{figure*}

The solution of the NP equation requires a relation between $c_{i}(\mathbf{r})$ and $\mu_{i}(\mathbf{r})$.
Here we use the LEMC simulation method that is an adaptation of the GCMC technique to a non-equilibrium situation.
We divide the computation domain of the NP system into elementary cells,  $\mathcal{D}^{\alpha}$, and use different $\mu_{i}^{\alpha}$ values in each volume element.
Insertions/deletions of ions are attempted into/from these volume elements with equal probability.
These trials are accepted or refused on the basis of the Metropolis algorithm \cite{metropolis}.
The acceptance probability contains the volume of the elementary cell, $V^{\alpha}$, the number of ionic species in that volume before insertion/deletion, $N_{i}^{\alpha}$, the local electrochemical potential, $\mu_{i}^{\alpha}$, and the energy change.
The energy includes every interaction from the whole simulation cell, not only from subvolume $\mathcal{D}^{\alpha}$.
The result of the LEMC simulation is the concentration in every volume element, $c_{i}^{\alpha}$.

The whole system is solved in an iterative way by adjusting the electrochemical potential profile ($\mu_{i}^{\alpha}$) in each iteration until conservation of mass ($\nabla \cdot \mathbf{j}_{i}(\mathbf{r})=0$) is satisfied.
In practice, the continuity equation is integrated for a volume element, $\mathcal{B}^{\alpha}$, and converted into a surface integral over $\mathcal{S}^{\alpha}$ on the basis of the Gauss-Ostogradsky theorem:
\begin{equation}
 0 = \int_{\mathcal{B}^{\alpha}}\nabla\cdot\mathbf{j}_{i}(\mathbf{r})\, dV = \oint_{\mathcal{S}^{\alpha}} \mathbf{j}_{i}(\mathbf{r}) \cdot \mathbf{n}(\mathbf{r})\,da ,
 \label{eq:cont2}
\end{equation}
where $\mathcal{B}^{\alpha}$ is bounded by $\mathcal{S}^{\alpha}$ and $\mathbf{n}(\mathbf{r})$ denotes the normal vector pointing outward at position $\mathbf{r}$ of the surface. 
Every $\mathcal{S}^{\alpha}$ surface is divided into $\mathcal{S}^{\alpha\beta}$ elements that separate adjacent cells, $\mathcal{B}^{\alpha}$ and $\mathcal{B}^{\beta}$.
Assuming that concentrations, electrochemical potentials, diffusion coefficients, and flux components are constant on a surface element superscripted by $\alpha\beta$, we can turn the integral into a sum:
\begin{equation}
 0 = \sum_{\beta, \mathcal{S}^{\alpha\beta}\in\mathcal{S}^{\alpha}} {\mathbf{j}}_{i}^{\alpha\beta} \cdot \mathbf{n}^{\alpha\beta}a^{\alpha\beta} .
 \label{eq:int-cont}
\end{equation}
Replacing the NP equation for $\mathbf{j}_{i}^{\alpha\beta}$ as $\mathbf{j}_{i}^{\alpha\beta}=-(1/kT)D_{i}^{\alpha\beta}c_{i}^{\alpha\beta}\nabla \mu_{i}^{\alpha\beta}$, we obtain a system of linear euation for $\mu_{i}^{\alpha\beta}$ for the next iteration using $c_{i}^{\alpha\beta}$ from the previous iteration.

Thus, the NP and the continuity equations together with LEMC form a self-consistent system.
The advantage over the widely used Poisson-Nernst-Planck theory is that the statistical mechanical component is an accurate particle simulation method instead of  an approximate mean--field theory (Poisson-Boltzmann) \cite{matejczyk_jcp_2017,madai_pccp_2018,valisko_jcp_2019}. 
Details are found in earlier papers  \cite{boda_jctc_2012,boda_jml_2014}.

The computational domain is a closed system (no periodic boundary conditions are applied).
We apply the boundary conditions that the concentrations of the ionic species and the electrical potential are constant on the boundary of our cell that is a cylinder in this study. 
Different values may be applied on the two sides of the membrane on the surfaces of the two half-cylinders as described in our previous studies \cite{boda_jctc_2012,boda_jml_2014}.
Since the boundary conditions are fixed, the driving force of the transport is maintained, so the transport is steady state. 

The length and radius of the cylindrical simulation cell were $300$ nm and $90$ nm, respectively.
Ion numbers depended on concentrations.
Typically, $80$ iterations were performed with $50,000$ LEMC steps in each iteration. 
Half of the LEMC steps were ion insertions and deletions ($50-50$ \%), while the other half were ion displacements.
This sampling was sufficient to get current data with error bars within the size of the symbols in the figures.

\section{Results}
\label{sec:results}

First, we report current-voltage ($I{-}U$) curves that are generally the raw outputs of experiments.
Fig.\ \ref{fig:fig2} show $I{-}U$ curves for various geometries, analyte ion valences ($z_{\mathrm{X}}=1$ and $2$) and concentrations ($c_{\mathrm{X}}=0$, $10^{-5}$, and $10^{-3}$ M).
The concentration of the background electrolyte (KCl) is $c_{\mathrm{KCl}}=0.01$ M.
Because the absolute values of the currents are different at positive (ON state) and negative (OFF state) voltages, the pore rectifies the ionic current. 
Also, both the current and the rectification (as we will see later) are sensitive to the value of the analyte concentration.
This fact makes this model nanopore applicable as a sensor.

\begin{figure*}[t]
	\begin{center}
		\includegraphics*[width=0.7\textwidth]{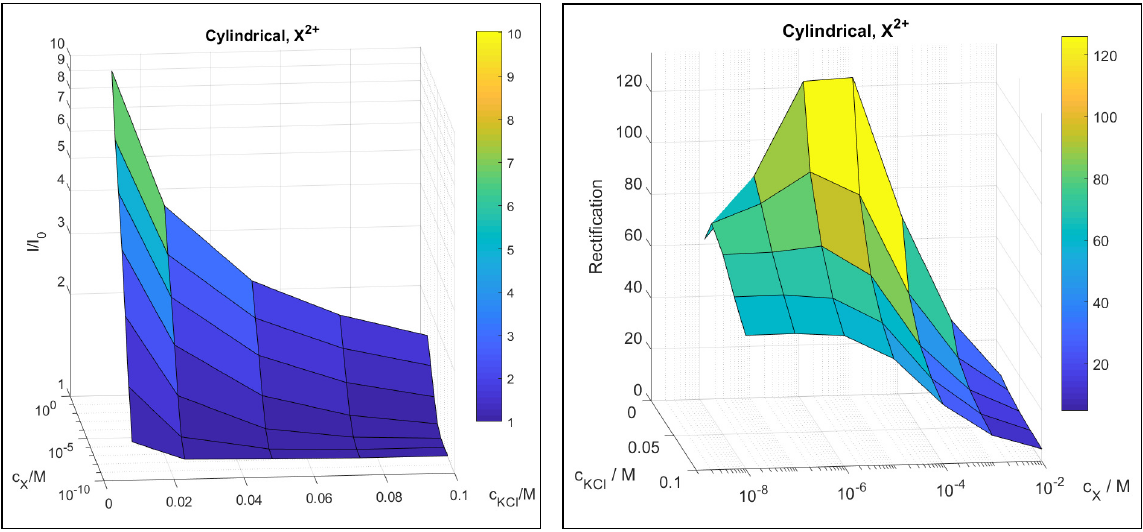}
		\caption{Dependence of the two device functions ($I/I_{0}$ in the left panel and rectification in the right panel) on $c_{\mathrm{X}}$ and $c_{\mathrm{KCl}}$ for the cylindrical pore and divalent analyte ions.  
		}
		\label{fig:fig4}
	\end{center}
\end{figure*}

Because the shapes of the $I{-}U$ curves are similar for every case, we can characterize the ON and OFF states by two chosen voltages, $\pm 200$ mV as in our previous studies \cite{hato_pccp_2017,matejczyk_jcp_2017,madai_jml_2019a,valisko_jcp_2019}.
Therefore, we define our device functions as the ON--state relative current $I^{\mathrm{ON}}/I^{\mathrm{ON}}_{0}$ normalized by the value at $c_{\mathrm{X}}=0$ and rectification $|I^{\mathrm{ON}}/I^{\mathrm{OFF}}|=|I(200\,\mathrm{mV})/I(-200\,\mathrm{mV})|$.
It is possible to define a rectification value relative to the $c_{\mathrm{X}}=0$ case: 
\begin{equation}
 \dfrac{|I^{\mathrm{ON}}/I^{\mathrm{OFF}}|}{|I_{0}^{\mathrm{ON}}/I_{0}^{\mathrm{OFF}}|}=
 \dfrac{ I^{\mathrm{ON}}/I_{0}^{\mathrm{ON}}}{I^{\mathrm{OFF}}/I_{0}^{\mathrm{OFF}}}.
\end{equation} 
The resulting formula is the ratio of the ON-- and OFF--state relative currents.
Once we have the $I{-}U$ curves for a range of $c_{\mathrm{X}}$ values, any of these quantities can be calculated.

In this study, we stay with the ON--state relative current (for which we will use the notation $I/I_{0}$ for briefness from now on) and the rectification because both are dimensionless quantities appropriate for our purposes.  
Thus, the basic results of our simulations are the device functions (relative current and rectification) plotted against the analyte concentration for a given set of the other parameters.
We call these curves calibration curves.

To understand the mechanisms behind the dual-response functioning of the nanopore, we examine the model by changing different parameters ($c_{\mathrm{KCl}}$, $z_{\mathrm{X}}$, $R_{\mathrm{X}}$, and pore geometry) and by plotting the calibration curves for various sets of parameters.
We generally fix all parameters and change a chosen one in order to be able to follow the trends. 

Fig.\ \ref{fig:fig3} shows the calibration curves for different values of the KCl concentration. 
The relative current is a monotonically increasing function of $c_{\mathrm{X}}$ for smaller KCl concentrations.
This device function, therefore, can be used for calibration if $c_{\mathrm{KCl}}$ is small.
With increasing KCl concentrations, the sensitivity of the relative current decreases. 
In some cases, even a minimum can be observed.

There is a decreasing trend in the rectification (see insets), but a maximum can be observed for divalent analyte ions, X$^{2+}$ (C and D panels), at smaller KCl concentrations.
The overall behavior is very similar for the cylindrical and conical geometries (left vs.\ right panels).

\begin{figure*}[t]
	\begin{center}
			\includegraphics*[width=0.7\textwidth]{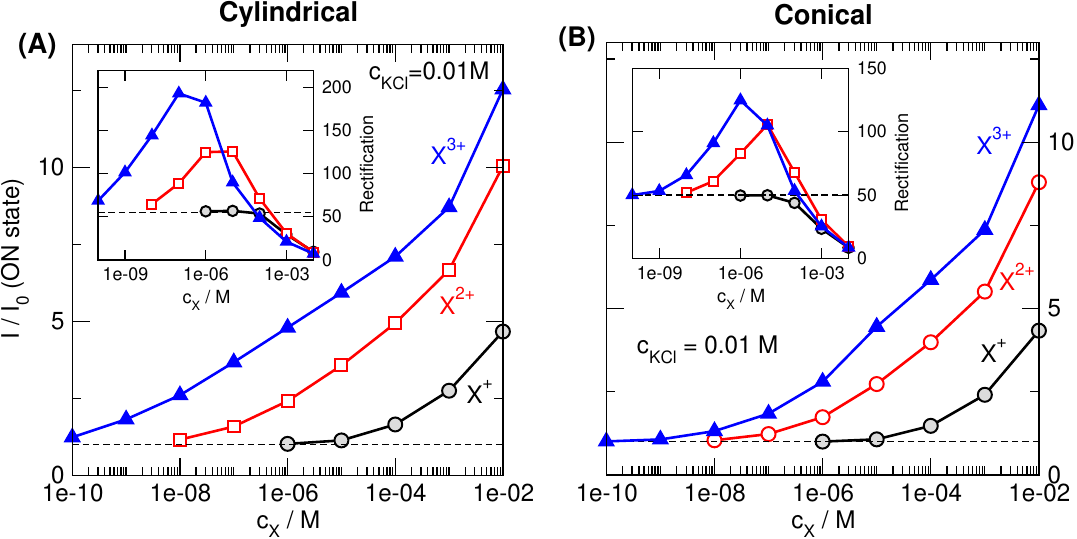}
		\caption{Relative currents ($I/I_{0}$) in the ON state (200 mV) as functions of analyte concentration ($c_{\mathrm{X}}$) for monovalent (black circles), divalent (red squares), and trivalent (blue triangles) analyte ions. 
		The insets show the rectification  as functions of $c_{\mathrm{X}}$.
		Left (A) and right (B) panels refer to the cylindrical and conical geometries, respectively.
		The concentration of the background electrolyte is $c_{\mathrm{KCl}}=0.01$ M.
		}
		\label{fig:fig5}
	\end{center}
\end{figure*}

The relation of the $c_{\mathrm{X}}$ and $c_{\mathrm{KCl}}$ concentrations is obviously important.
Fig.\ \ref{fig:fig4} shows the device functions, current ratio (panel A) and rectification (panel B), as functions of both concentrations for the divalent analyte ions.
The two panels of this figure can be looked at as two ``calibration surfaces'' for the two device functions as functions of the two concentrations in question.
How the balance of these concentrations determines device functions will be studied in the Discussions (section\ \ref{sec:discussion}).
The figure demonstrates the dual--response nature of our device.

The main conclusion of Figs.\ \ref{fig:fig3} and \ref{fig:fig4} is that a small KCl concentration is more appropriate if we want to use $I/I_{0}$ as our device function in accordance with experiments \cite{tang_sab_2019}. 
There is a robust increase in the current that is chiefly anion current without ``cation leakage'' (see later).
If we want to use the  rectification as our device function, the choice is not so obvious.
The $c_{\mathrm{X}}$--dependence of the rectification is monotonic at larger KCl concentrations that can be handy in certain situation.
The maximum, on the other hand can be a hallmark of multivalent ions that can be helpful if we want to determine the charge of an analyte.

Therefore, we fix KCl concentration at $c_{\mathrm{KCl}}=0.01$ M and plot results for analyte ions with different valences in Fig.\ \ref{fig:fig5}.
Currents are larger for analytes with larger valences because they attract more anions into the right region.
It was already apparent from Fig.\ \ref{fig:fig3}, but this figure shows it explicitly.
The behavior of rectification is even more interesting.
There is no maximum for monovalent analyte ions.
The maximum appears in the case of divalent analyte ions, while it is shifted to the left (into the direction of smaller $c_{\mathrm{X}}$ values) in the case of trivalent analyte ions.

This fact can help to identify the valence of the X$^{z_{\mathrm{X}}}$ ion by measuring the rectification for a series of dilution. 
The analyte concentration as a function of rectification is not a one-value function in this case, unfortunately, but the value of $I/I_{0}$, for example, can help us in the decision whether we are in the increasing or the decreasing branch of the rectification curve.

\begin{figure*}[t]
	\begin{center}
		\includegraphics*[width=0.75\textwidth]{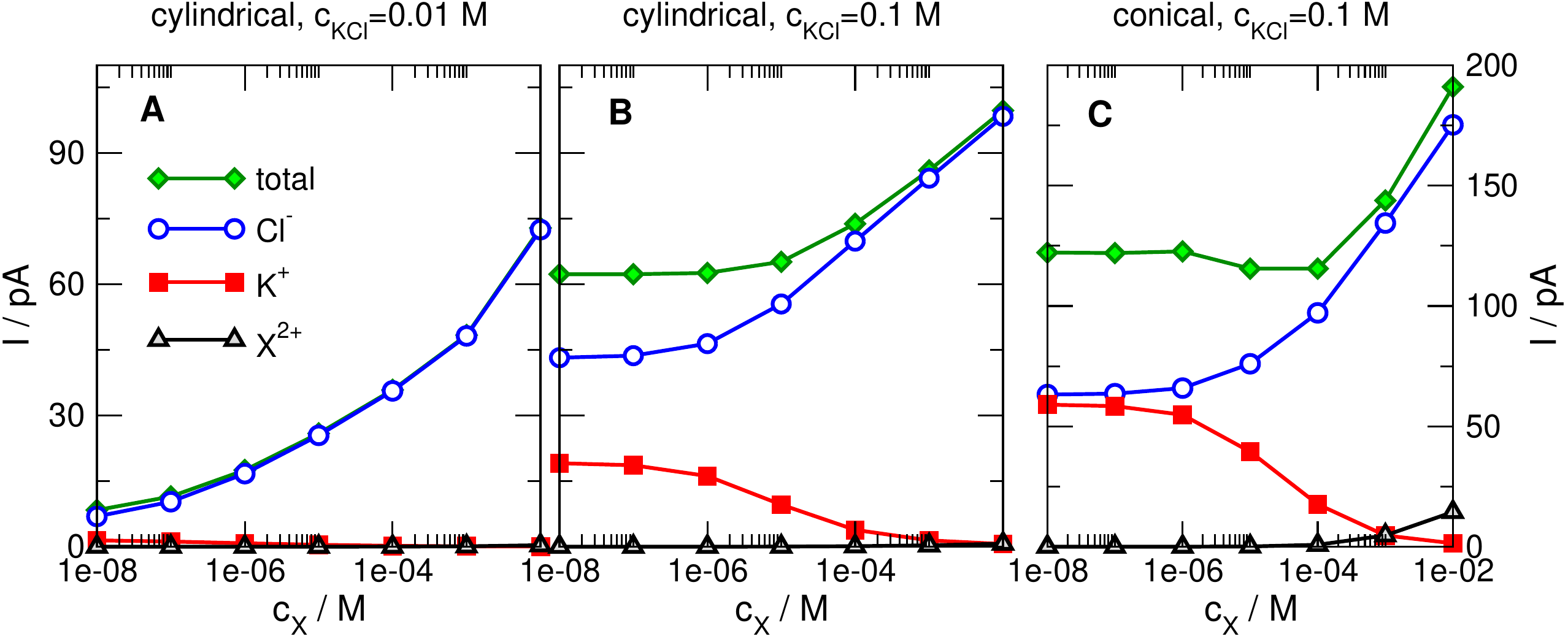}
		\caption{Individual ionic currents carried by Cl$^{-}$ (blue circles), K$^{+}$ (red squares), and X$^{2+}$ (black triangles) as functions of $c_{\mathrm{X}}$ in the ON state.
		The total currents are shown with green diamonds.
		Panels A and B refer to background electrolyte concentrations $c_{\mathrm{KCl}}=0.01$ and $0.1$ M, respectively, for the cylindrical geometry (these two panels have the same scale on the abscissa).
		Panel C refers to background electrolyte concentration $0.1$ M for the conical geometry. 
		The analyte ions are divalent.
		}
		\label{fig:fig6}
	\end{center}
\end{figure*}

\section{Discussion}
\label{sec:discussion}

In this section we look behind the curtain  and analyze the currents carried by the individual ionic species and their concentration profiles.
One important advantage of the NP+LEMC technique is that it straightforwardly provides concentration, electrochemical potential, electrical potential, and flux density profiles from which the ionic concentration profiles are the most important, because they are the determining factors of the current. 

\begin{figure*}[t]
	\begin{center}
		\includegraphics*[width=0.68\textwidth]{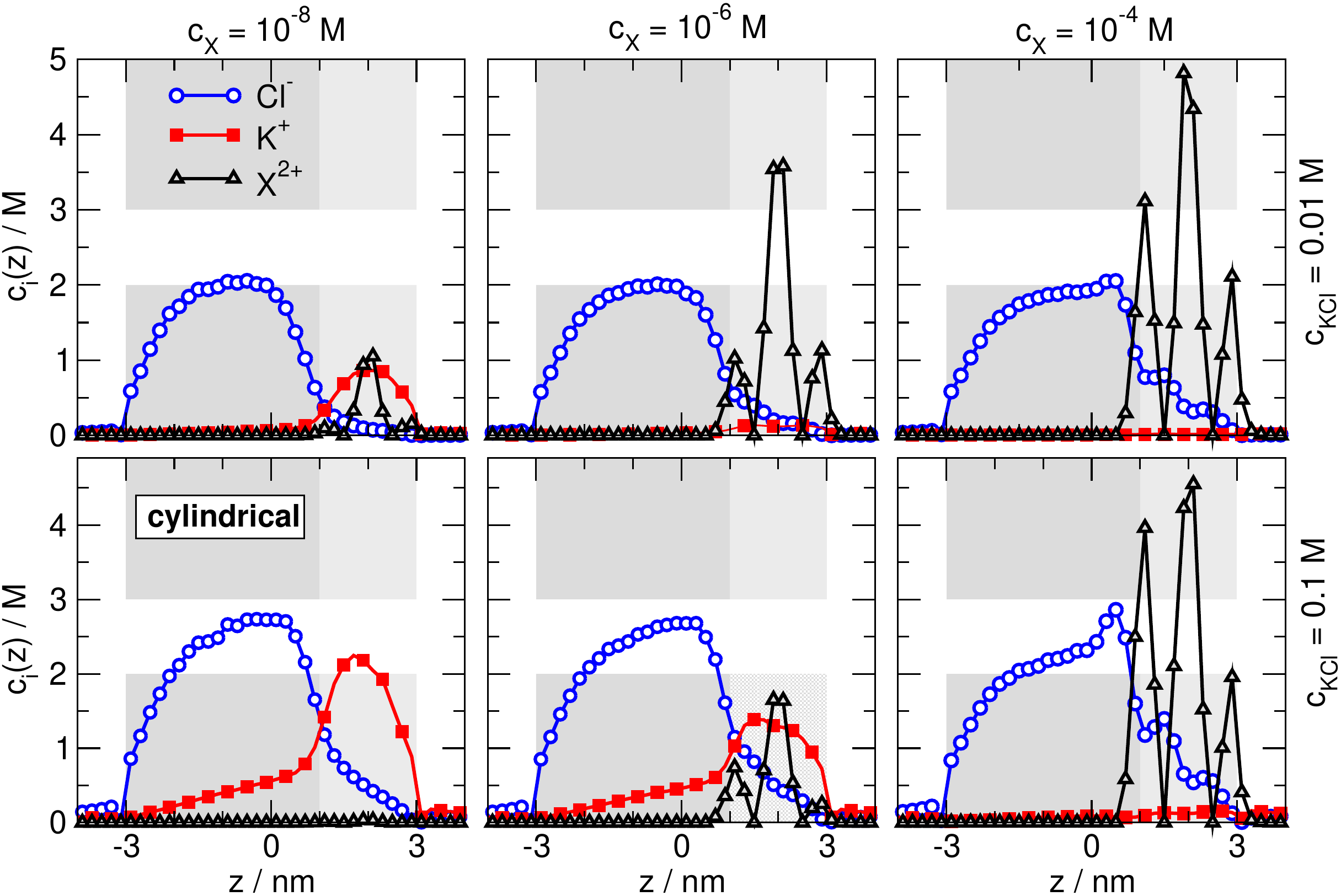}
		\caption{Axial concentration profiles for analyte concentrations $c_{\mathrm{X}}=10^{-8}$ M (left column), $10^{-6}$ M (middle column), and $10^{-4}$ M (right column).
		Top and bottom rows refer to background electrolyte concentrations $c_{\mathrm{KCl}}=0.01$ M and $0.1$ M, respectively. 
		The figure shows results for the cylindrical geometry and divalent analyte ions in the ON state.
		}
		\label{fig:fig7}
	\end{center}
\end{figure*}

If the concentration of a given ionic species is very small in a given region of the pore along the $z$-axis, then this region behaves as a large-resistance element of the regions imagined as resistors connected in series.
This large-resistance region, called depletion zone, tend to determine the resistance of the whole device especially in the case of bipolar nanopores whose device function   is determined by the appearance of depletion zones controlled by the sign of the voltage and the polarity of the surface charge distribution \cite{hato_pccp_2017}. 

Let us first consider the question why the current behaves so differently at small and large KCl concentrations.
Fig.\ \ref{fig:fig6} shows the currents carried by individual ions together with the total current for $c_{\mathrm{KCl}}=0.01$ M (panel A) and $c_{\mathrm{KCl}}=0.1$ M (panel B) for the cylindrical geometry.
Panel C refers to $c_{\mathrm{KCl}}=0.1$ M for the conical geometry. 
The case of divalent analyte ions is considered.
The Cl$^{-}$ current monotonically increases with $c_{\mathrm{X}}$ in all cases.
The main difference is that the K$^{+}$ current is practically absent in the case of $c_{\mathrm{KCl}}=0.01$ M (Fig.\ \ref{fig:fig6}A), while it is considerable in the case of $c_{\mathrm{KCl}}=0.1$ M (Figs.\ \ref{fig:fig6}B and C).
This K$^{+}$ leakage is even larger in the conical geometry (Fig.\ \ref{fig:fig6}C).

K$^{+}$ leakage decreases with increasing $c_{\mathrm{X}}$, therefore, it counterbalances the increase of Cl$^{-}$ current.
The total current, therefore, is less sensitive to the analyte concentration at high $c_{\mathrm{KCl}}$.
Consequently, the relative current, $I/I_{0}$, can be used as a device function efficiently only if the K$^{+}$  leakage is absent, namely, at low KCl concentrations.

The concentration profiles of Cl$^{-}$, K$^{+}$, and X$^{2+}$ for three selected $c_{\mathrm{X}}$ values of Figs.\ \ref{fig:fig6}A and B are plotted in Fig.\ \ref{fig:fig7}.
As the X$^{2+}$ ions accumulate in the pore with increasing $c_{\mathrm{X}}$ (from left to right), they squeeze K$^{+}$ ions out of the pore, especially out of the right (binding) region.
The major difference between the $c_{\mathrm{KCl}}=0.01$ and $0.1$ M cases is that there are generally less K$^{+}$ and Cl$^{-}$ ions in the nanopore at small $c_{\mathrm{KCl}}$.
Because the K$^{+}$ ion is the coion, being less of it means that it is depleted enough so that K$^{+}$ current is negligible (Fig.\ \ref{fig:fig6}A).
For $c_{\mathrm{KCl}}=0.1$ M, K$^{+}$ concentration is large in the pore and there is considerable K$^{+}$ leakage (Fig.\ \ref{fig:fig6}B--C).

\begin{figure}[t]
	\begin{center}
		\includegraphics*[width=0.35\textwidth]{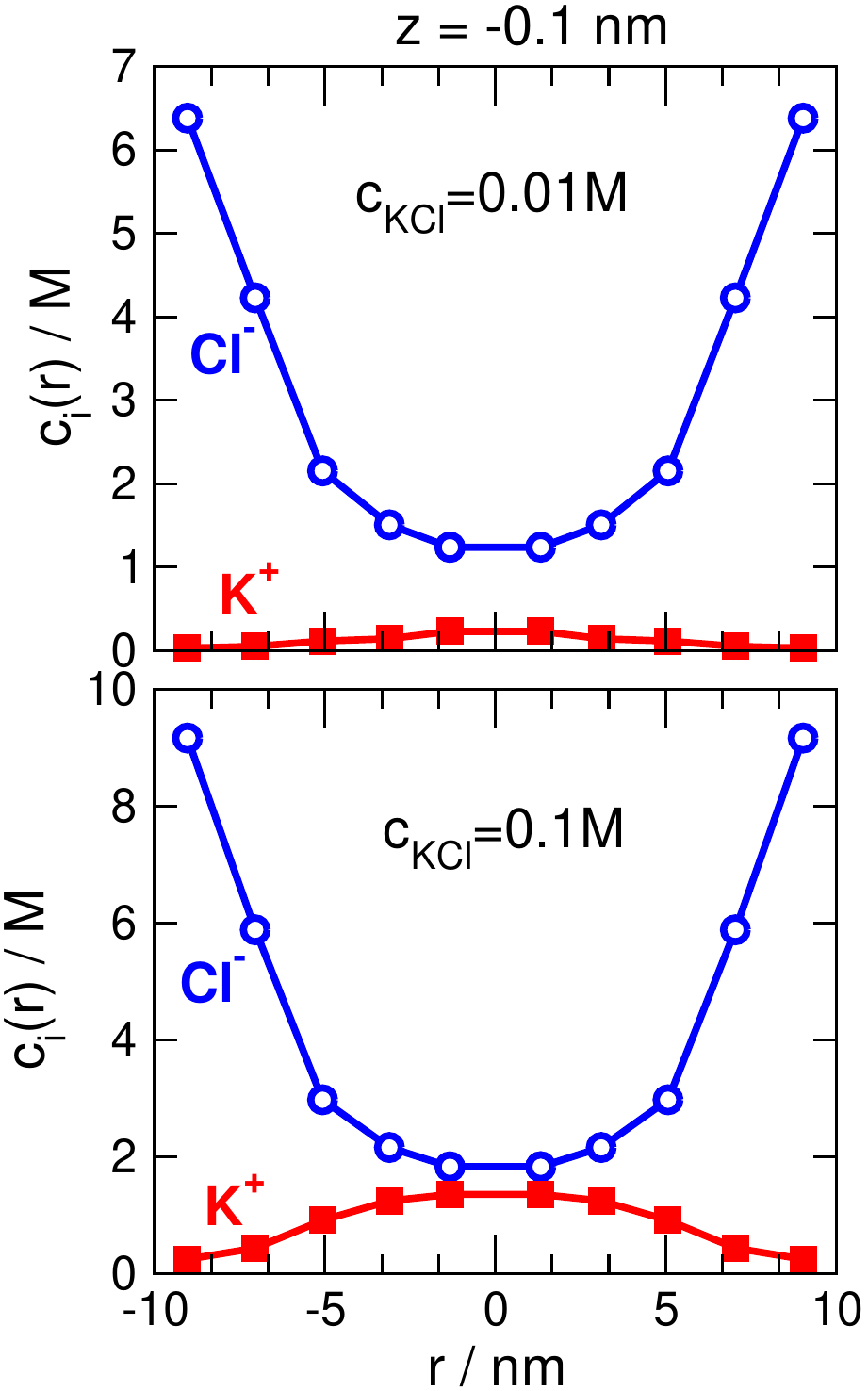}
		\caption{
		Radial concentration profiles of Cl$^{-}$ and K$^{+}$ at $z=-0.1$ nm, a selected position in the left region, for $c_{\mathrm{KCl}}=0.01$ M (top) and $0.1$ M (bottom) for the cylindrical geometry, $c_{\mathrm{X}}=10^{-8}$ M, and X$^{2+}$ (left column of Fig.\ \ref{fig:fig7}).
		}
		\label{fig:fig8}
	\end{center}
\end{figure}

The difference between the smaller and larger KCl concentrations can also be depicted from looking at the radial profiles.
Nanopores are distinguished from micropores by the fact that the characteristic screening length (Debye length) of the electrolyte is measurable to the radius of the pore. 
In this case, the double layers that are formed near the charged wall of the pore in the radial dimension extend into the central region of the pore (see Fig.\ \ref{fig:fig8}, where the radial profiles are plotted for a selected $z$ coordinate in the left region of the pore).
Smaller bulk concentration results in larger Debye-length, so the double layers overlap near the centerline ($r=0$) as shown by the top panel of Fig.\ \ref{fig:fig8}. 
In this case, a charge neutral bulk-like region does not form along the centerline. 
The result is that a depletion zone of the coions (K$^{+}$, in this case) appear.
Formation of this depletion zone prevents K$^{+}$ leakage.

\begin{figure}[t!]
	\begin{center}
		\includegraphics*[width=0.35\textwidth]{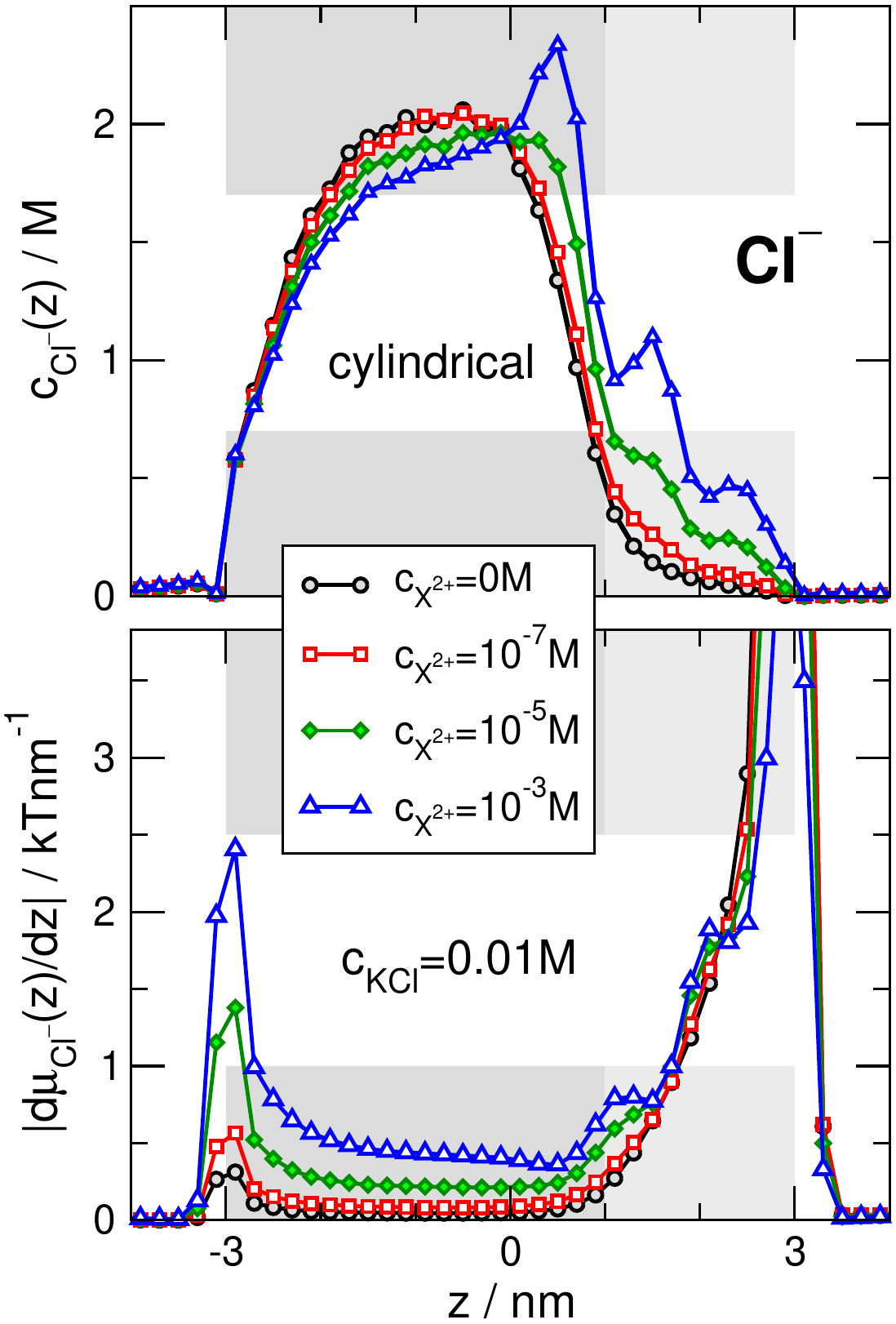}
		\caption{Axial concentration profiles (top) and driving forces (the $z$-derivatives of the axial, cross-section averaged electrochemical potential profiles; bottom panel) for the anion (Cl$^{-}$) for different analyte concentrations.
		The figure shows results for the cylindrical geometry, divalent analyte ions, and $c_{\mathrm{KCl}}=0.01$ M.
		}
		\label{fig:fig9}
	\end{center}
\end{figure}

The basis of sensing is the increasing Cl$^{-}$ current with increasing analyte concentration.
The increase of Cl$^{-}$ current is fundamentally due to the increase of Cl$^{-}$ concentration in the right region as more Cl$^{-}$ ions are attracted by the accumulating analyte ions.
Fig.\ \ref{fig:fig7} does not show the increase of Cl$^{-}$ concentrations very clearly.
Furthermore, not only the concentration, but also the driving force ($\nabla  \mu_{i}$) can play an important role in the behavior of the Cl$^{-}$ current.
Therefore, we plot Cl$^{-}$ concentration profiles (top) and the $-\mathrm{d} \mu_{\mathrm{Cl}^{-}}(z)/\mathrm{d}z$ profiles (bottom) in Fig.\ \ref{fig:fig9}. 
The $\mathrm{d} \mu_{\mathrm{Cl}^{-}}(z)/\mathrm{d}z$ derivative corresponds to the $z$-component of $\nabla \mu_{\mathrm{Cl}^{-}}(\mathbf{r})$ if the electrochemical potential is approximately constant in the radial dimension, which is the case here (results not shown).

Top panel of Fig.\ \ref{fig:fig9} shows that more Cl$^{-}$ ions accumulate in the right region as $c_{\mathrm{X}}$ increases.
At the same time, however, Cl$^{-}$ concentration decreases in the left region.
Eventually, the current of Cl$^{-}$ increases because the driving force ($|\mathrm{d} \mu_{\mathrm{Cl}^{-}}(z)/\mathrm{d}z|$, bottom panel of Fig.\ \ref{fig:fig9}) increases in the left region, while it is relatively unchanged in the right region.
This behavior is the result of the self-consistent solution of the NP+LEMC system.

Figs.\ \ref{fig:fig3}--\ref{fig:fig5} show that rectification has a maximum as a function of $c_{\mathrm{X}}$ for small values of $c_{\mathrm{KCl}}$ and multivalent analyte ions ($z_{\mathrm{X}}=2$ and $3$).
To analyze the mechanism behind this behavior, we choose one case ($z_{\mathrm{X}}=3$, $c_{\mathrm{KCl}}=0.01$ M, and cylindrical geometry; see the inset of Fig.\ \ref{fig:fig5}A) and show profiles for that case. 

\begin{figure*}[t]
	\begin{center}
		\includegraphics*[width=0.65\textwidth]{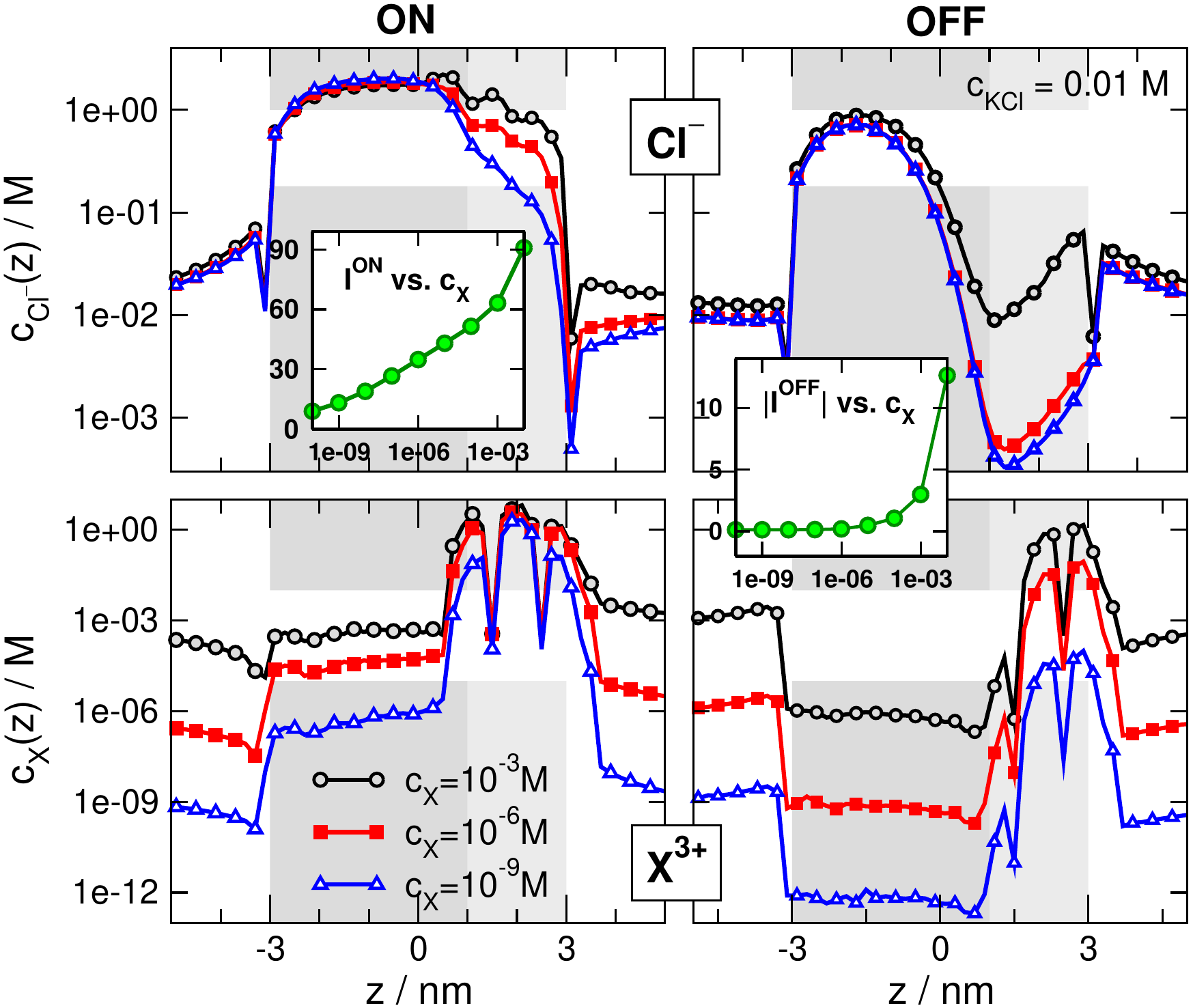}
		\caption{Concentration profiles of Cl$^{-}$ (top row) and X$^{3+}$ (bottom row) for three selected analyte concentrations $c_{\mathrm{X}}=10^{-9}$, $10^{-6}$, and $10^{-3}$ M.
		Left and right columns show the results for the ON and OFF states, respectively.
		The $c_{\mathrm{X}}$ concentrations are chosen such a way that they represent three characteristic points of the rectifications vs.\ $c_{\mathrm{X}}$ curve (see blue triangles in the inset of Fig.\ \ref{fig:fig5}A) with the $10^{-6}$ M point being close to the maximum.
		The insets show the current vs. $c_{\mathrm{X}}$ curves.
		Dividing the ON curve with the OFF curve, we get the rectification curve of Fig.\ \ref{fig:fig5}A.
		}
		\label{fig:fig10}
	\end{center}
\end{figure*}

Fig.\ \ref{fig:fig10} shows Cl$^{-}$ and X$^{3+}$ profiles for  three selected analyte concentrations $c_{\mathrm{X}}=10^{-9}$, $10^{-6}$, and $10^{-3}$ M.
They have been chosen so that they represent three characteristic points of the rectification vs.\ $c_{\mathrm{X}}$ curve with the $10^{-6}$ M point being close to the maximum.
The left column shows the results for the ON state, while the right column for the OFF state.
The maximum in the rectification vs.\ $c_{\mathrm{X}}$ appears because of the behavior of the $I^{\mathrm{ON}}(c_{\mathrm{X}})$ and $|I^{\mathrm{OFF}}(c_{\mathrm{X}})|$ curves (they are shown in the insets of Fig.\ \ref{fig:fig10}).
For lower $c_{\mathrm{X}}$ concentrations (below $10^{-6}$ M) the current increases faster in the ON state than in the OFF state.
In this $c_{\mathrm{X}}$ regime, therefore, rectification increases. 
Above $c_{\mathrm{X}}=10^{-6}$ M, however, the OFF--current starts to increase heavily that causes a decrease in the rectification.

The explanation of this behavior of $I^{\mathrm{ON}}(c_{\mathrm{X}})$ and $I^{\mathrm{OFF}}(c_{\mathrm{X}})$ can be depicted from the profiles.
In the ON state, the applied field favors the accumulation of X$^{3+}$ ions in the pore.
There are enough analyte ions bound, so they can have a considerable effect on the Cl$^{-}$ ions.
As $c_{\mathrm{X}}$ is increased from $10^{-9}$ M to $10^{-6}$ M, there is a corresponding increase in Cl$^{-}$ concentration on the right hand side.

In the OFF state, however, the applied field hinders accumulation of X$^{3+}$ ions in the pore, so the concentration of X$^{3+}$ ions remains small even at $c_{\mathrm{X}}=10^{-6}$ M. 
Thus, as we increase $c _{\mathrm{X}}$ from $10^{-9}$ M to $10^{-6}$ M, we do not gain much increase in Cl$^{-}$, because there are no enough X$^{3+}$ ions to attract them in.
The OFF--current, therefore, does not change much in this $c_{\mathrm{X}}$ regime. 

Above $c_{\mathrm{X}}=10^{-6}$ M, however, there are enough X$^{3+}$ ions in the pore so that the Cl$^{-}$ ions ``feel'' their presence.
In this $c_{\mathrm{X}}$ regime, excess accumulation of Cl$^{-}$ ions on the right hand side shoots off that creates the downhill side of the rectification vs.\ $c_{\mathrm{X}}$ function.

Figs.\ \ref{fig:fig2}. \ref{fig:fig3}, and \ref{fig:fig5} show the device functions for both the cylindrical and conical geometries.
There is no considerable difference between these geometries regarding the behavior of currents and rectification.
Currents are somewhat larger in the conical pores due to their larger cross sections.
The overall qualitative behavior, however, is the same.
This is comforting, because technically it is easier to manufacture a conical nanopore from PET with the etching technique. 

The explanation of the qualitatively similar behavior is that the important things happen in the right hand region of the pore, where binding of the analyte ions and excess accumulation of Cl$^{-}$ ions occur.
This region is the tip of the conical nanopore similar in size to the cylindrical pore.
The left region is a buffer region whose job is to determine the main charge carrier (Cl$^{-}$) and to exclude the coion (K$^{+}$).

\begin{figure*}[t]
	\begin{center}
		\includegraphics*[width=0.65\textwidth]{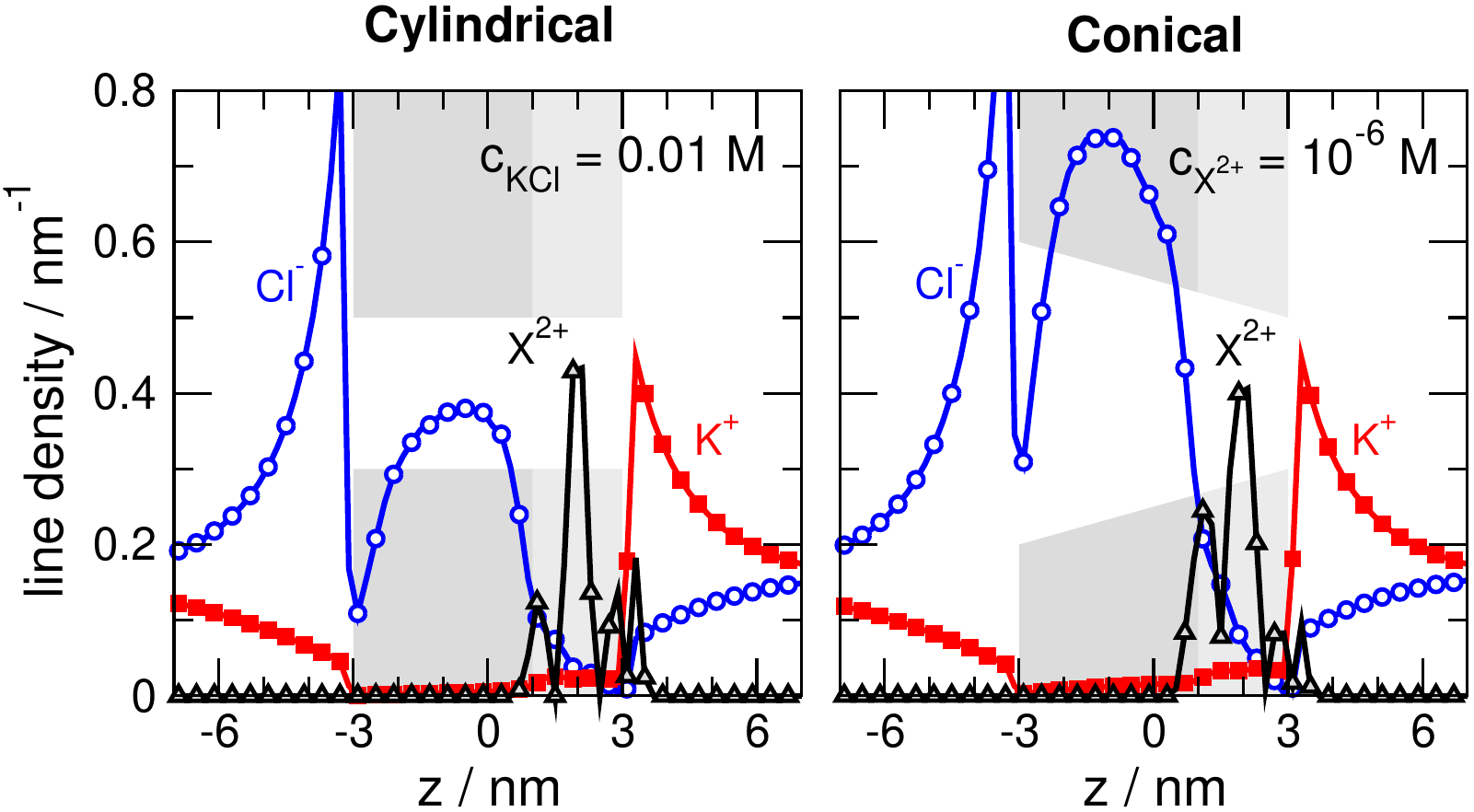}
		\caption{Line densities of Cl$^{-}$, K$^{+}$, and X$^{2+}$ for $c_{\mathrm{X}}=10^{-6}$ M and $c_{\mathrm{KCl}}=0.01$ M. 
		Line density is defined as $n_{i}(z)=\langle N(z)\rangle /\Delta z$ (the average number of ions in a slab of width $\Delta z$ divided by $\Delta z$).
		Left and right panels show the results for the cylindrical and conical geometry, respectively.
		}
		\label{fig:fig11}
	\end{center}
\end{figure*}

Fig.\ \ref{fig:fig11} shows a typical case for divalent analyte ions, $c_{\mathrm{X}}=10^{-6}$ M, and $c_{\mathrm{KCl}}=0.01$ M.
The figure shows the line density that is defined as the average number of ions in a slab of width $\Delta z$ divided by $\Delta z$: $n_{i}(z)=\langle N(z)\rangle /\Delta z$.
This profile is not normalized with the cross section as opposed to the concentration profile. 
Therefore, line density is related to the actual number of ions in a certain segment of the pore instead of their density.
The relation between the two functions is
\begin{equation}
 c_{i}(z) = \dfrac{n_{i}(z)}{N_{\mathrm{A}}10^{-24}A(z)},
 \label{ni}
\end{equation} 
where $N_{\mathrm{A}}$ is Avogadro's number and $A(z)$ is the cross section of the pore at $z$ (outside the pore, $A(z)$ is the cross section of the cylindrical simulation cell).
Eq.\ \ref{ni} holds if we measure $n_{i}$ in nm, $A_{i}$ in nm$^{2}$, and $c_{i}$ in mol/dm$^{3}$.

Fig.\ \ref{fig:fig11} shows that there are more ions (Cl$^{-}$ and X$^{2+}$) accumulated in the conical pore as we proceed from right to left towards the wide opening of the pore. 
This increases the magnitudes of currents (see Fig.\ \ref{fig:fig2}), but we use the relative currents, $I/I_{0}$, as device function.
So, if the difference between, let us say, the $c_{X}=10^{-6}$ M and $c_{\mathrm{X}}=0$ M cases are the same, the $I/I_{0}$ device function does not change considerably. 
We have seen in Figs.\ \ref{fig:fig3} and \ref{fig:fig5} that this is the case.

Rectification is also a normalized quantity similarly to $I/I_{0}$. 
So, if the difference between the ON and OFF states is similar in the two geometries, then the other device function, rectification, should not be sensitive to the geometry either.
We have seen in Figs.\ \ref{fig:fig3} and \ref{fig:fig5} that this is the case.

The main disadvantage of using a conical pore is that we may get K$^{+}$ leakage.
This, however, can be prevented by using an appropriately chosen concentration.
As we have seen in our previous study \cite{madai_pccp_2018} for a nanopore transistor, the relevant parameter is $R_{\mathrm{pore}}/\lambda_{\mathrm{D}}$, where $R_{\mathrm{pore}}$ is the pore radius somewhere along the pore and $\lambda_{\mathrm{D}}$ is the Debye length. 
In that paper \cite{madai_pccp_2018} we showed that device functions scale with the $R_{\mathrm{pore}}/\lambda_{\mathrm{D}}$ parameter.
It means that we get the same device behavior at a different $R_{\mathrm{pore}}$ if we choose a concentration that produces the same  $R_{\mathrm{pore}}/\lambda_{\mathrm{D}}$ ratio.
Specifically, we get the same device function for a narrow pore with large concentration and for a wide pore with low concentration as soon as the $R_{\mathrm{pore}}/\lambda_{\mathrm{D}}$ parameter is the same.

Fig.\ \ref{fig:fig11} also shows the double layers formed at the membrane on the left and right hand sides. 
These double layers are also seen in the concentration profiles (Fig.\ \ref{fig:fig10}, for example), but plotting line densities magnifies them because the baths on the two sides of the membrane  in the simulation cell are large.
Fig.\ \ref{fig:fig11} shows results for the ON state (positive electrical potential on the right hand side), so there is an excess K$^{+}$ accumulation in the right hand side double layer, while there is an excess Cl$^{-}$ accumulation in the left hand side double layer.

\begin{figure*}[t]
	\begin{center}
		\includegraphics*[width=0.7\textwidth]{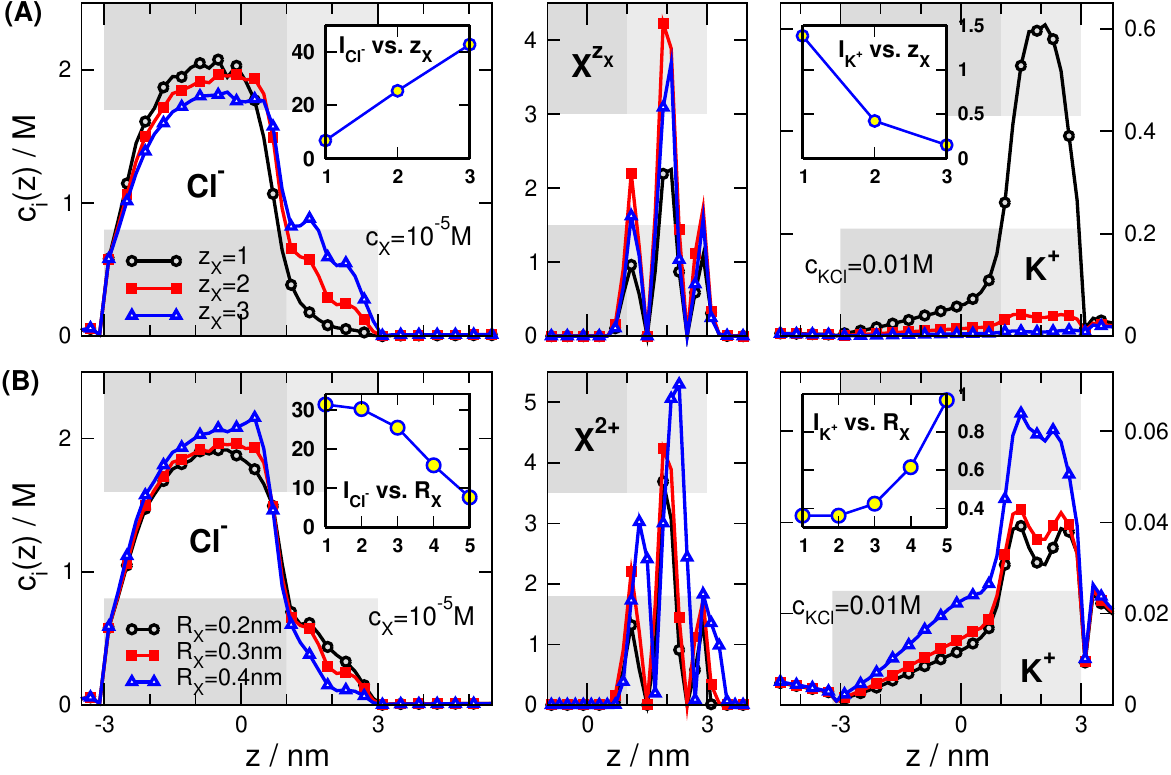}
		\caption{Concentration profiles of Cl$^{-}$ (left panel), X$^{z_{\mathrm{X}}}$ (middle panel), and K$^{+}$ (right panel)  (A) for varying analyte valences and (B) for varying analyte radii.
		The parameters are $c_{\mathrm{X}}=10^{-5}$ M, $c_{\mathrm{KCl}}=0.01$ M, cylindrical geometry, and ON state.
        The insets show the currents of the respective ionic species as functions of (A) $z_{\mathrm{X}}$ and (B) $R_{\mathrm{X}}$.
		}
		\label{fig:fig12}
	\end{center}
\end{figure*}

Finally, we discuss the effect of the properties of the analyte ions, valence and radius, on currents and concentration profiles.
Fig.\ \ref{fig:fig5} already showed that relative currents are larger for multivalent analyte ions, while the rectification shows an interesting maximum discussed earlier.
The increase of current as a function of $z_{\mathrm{X}}$ is analyzed in Fig.\ \ref{fig:fig12}A.
The insets show that Cl$^{-}$ current increases, while K$^{+}$ current decreases with increasing $z_{\mathrm{X}}$.
This behavior can be understood from the concentration profiles.

There is more X$^{2+}$ ions bound than X$^{+}$ ions due to the increased attraction by the right region's negative surface charge.
Furthermore, the charge carried by the X$^{2+}$ ions is twice as much carried by the X$^{+}$ ions.
Accordingly, more Cl$^{-}$ ions are attracted into the right region, while more K$^{+}$ ions are repulsed from the right region.
If we increase the valence further to $z_{\mathrm{X}}=3$, a saturation is observed: the analyte concentration is not increased further.
The charge carried by the X$^{3+}$ ions, however, is $1.5$ times larger than carried by the X$^{2+}$ ions, so a further increase in Cl$^{-}$ concentration and a further decrease in K$^{+}$ concentration is found.

The enhanced accumulation of anions casued by more strongly adsorbed multivalent cations resembles the phenomenon of charge inversion when multivalent cations overcharge the negative surface charge and they attract more anions into the pore causing reversal of ion selectivity \cite{he_jacs_2009} or electrokinetic mobility \cite{van_der_Heyden_prl_2006}.
In our case, however, the primary interaction attracting cations into the pore is the SW potential, while multivalent cations enhance this effect by their stronger electrostatic interactions.

When the radius of the analyte ions is changed (see Fig.\ \ref{fig:fig12}B), we observe a similar behavior: smaller analyte ions can find space easier in the  pore and they are attracted more strongly by the surface charge.
The concentration of the X$^{2+}$ ions in the pore, therefore, decreases with increasing $R_{\mathrm{X}}$ with a saturation at larger radii.
The response of Cl$^{-}$ and K$^{+}$ ions, however, is continuous.
Larger X$^{2+}$ ions can attract less Cl$^{-}$ ions, because the charge in the center of the larger X$^{2+}$ ions can not get  so close to the charge of the Cl$^{-}$ ions.
This means that the effective Coulomb interaction is weaker that is generally characterized by the $z_{\mathrm{X}}z_{\mathrm{Cl}^{-}}e^{2}/4\pi\epsilon_{0}\epsilon (R_{\mathrm{X}}+R_{\mathrm{Cl}^{-}})$ coupling constant.

Fig.\ \ref{fig:fig12} shows that our device can detect multivalent analyte ions more efficiently, while it can detect larger analyte ions less efficiently.
In reality, multiply charged ions tend to be larger, so these two effects tend to balance each other. 

\section{Summary}

This work shows that using a handful of interactions in a reduced model of a nanopore-based sensor, we can create a sensor model producing diverse behavior.
The interactions in question include long-range Coulombic forces acting between ions, pore charges, and applied field as well as short-range forces including hard-sphere and  hard-wall exclusion and the SW attraction by the binding sites.
The competition of all these forces produces a behavior that is complicated further by the asymmetric nature of the nanopore.

This balance can also be shifted by the presence of polarization charges that would arise at the dielectric boundaries for an inhomogeneous dielectric when $\epsilon$ is different in the electrolyte and in the membrane.
The interaction with surface charge would either be enhanced or weakened by induced charge depending on the value of the dielectric constant in the membrane.
The basic mechanism of sensing, however, would remain the same.
Although this model would be more realistic, its simulation would require much more computation time.

Pore asymmetry is created by the bipolar pattern of surface charges and by the asymmetric placement of the binding sites in the negatively charged region (asymmetry in pore shape seems to be less important).
We showed that due to the asymmetric structure of the pore, rectification appears as a device function in addition to the current in the ON state of the pore.
We showed that low concentration of the background electrolyte (KCl) is advantageous in preventing K$^{+}$ leakage and in producing a monotonically $c_{\mathrm{X}}$--dependent Cl$^{-}$ current that can serve as a calibration curve.
Rectification shows a maximum that appears  due to the competition between different forces as discussed in the main text.

The device can detect multivalent ions more efficiently especially if their size is smaller.
They can be measured in concentrations way below micromolar.
This indicates that the mechanism behind the $c_{\mathrm{X}}$--sensitive response of the device is appropriate to detect trace concentrations.
The basic reason of this capability is the small size of the nanopore that is comparable to the Debye length of the electrolyte so tuning depletion zones is possible by very small changes in the conditions.
Small changes in concentrations (from $c=10^{-6}$ to $10^{-4}$, for example) can result in a considerable change in the chemical potential ($kT\ln 100$ in the example), because it depends on $\ln c$.
That change in the chemical potential influences the energetic competition \cite{boda_jcp_2011_analyze} inside the pore. 
The balance of thermodynamic parameters and local interactions results in ionic currents that respond sensitively to changes in any of the competing effects.

Basically, the efficiency of the device is a consequence of the fact that the local electric field in the nanopore is sensitive to the binding of the analyte ions.
Changes in thermodynamic conditions in the bath, therefore, actuate changes in electric field locally in the nanopore, that, in turn, actuate changes in local ionic concentrations, and, therefore, in current.
Modeling and simulation of these intricate chains of events can help in designing and building more efficient nanopore sensors.

\section*{Acknowledgements}
\label{sec:ack}

We gratefully acknowledge  the financial support of the National Research, Development and Innovation Office -- NKFIH K124353. 
Present article was published in the frame of the project GINOP-2.3.2-15-2016-00053.

%

\end{document}